\pdfoutput=1

\documentclass[aps,twocolumn,groupedaddress,longbibliography]{revtex4-2}

\usepackage{amsmath}
\usepackage{siunitx}
\usepackage{graphicx}
\usepackage{physics}
\usepackage{xspace}
\usepackage{mathtools}

\usepackage{acronym}
\newacro{ATPa}[ATP]{adenosine triphosphate}
\newacro{ATP}[ATP]{adenosine triphosphate}
\newacro{ADP}[ADP]{adenosine diphosphate}
\newacro{TASAM}[TASAM]{totally asymmetric allosteric model}

\newcommand{\Fone}{F\textsubscript{1}\xspace}
\newcommand{\FoFone}{F\textsubscript{o}F\textsubscript{1}\xspace}
\newcommand{\Nex}{N_\mathrm{ex}}

\newcommand{\kb}{k_\mathrm{B}}

\newcommand{\ATP}{\mathrm{ATP}}
\newcommand{\ADP}{\mathrm{ADP}}
\newcommand{\Vmax}{V_\mathrm{max}}
\newcommand{\Km}{K_\mathrm{m}}
\newcommand{\Subs}{\mathrm{S}}
\newcommand{\iP}{\textrm{P}\textsubscript{i}\xspace}
\newcommand{\B}{\mathrm{b}}
\newcommand{\h}{\mathrm{h}}
\newcommand{\tdm}{\widetilde{\Delta\mu}}
\newcommand{\Weff}{W_\mathrm{eff}}

\begin{document}

\title{
Asymmetric enzyme kinetics of \Fone-ATPase
\\ induced by rotation-assisted substrate binding 
}

\author{Yohei Nakayama}
\affiliation{Department of Applied Physics, Graduate School of Engineering, Tohoku University, Aoba 6-6-05, Sendai 980-8579, Japan}
\author{Shoichi Toyabe}
\email{toyabe@tohoku.ac.jp}
\affiliation{Department of Applied Physics, Graduate School of Engineering, Tohoku University, Aoba 6-6-05, Sendai 980-8579, Japan}

\date{\today}

\begin{abstract}
We demonstrate asymmetric enzyme kinetics of a biomolecular motor \Fone-ATPase between synthesis and hydrolysis of \ac{ATPa}.
Our experiments show that ATP hydrolysis follows Michaelis-Menten kinetics, but ATP synthesis, which is an \Fone-ATPase's primary biological role, deviates from it. Specifically, the synthesis rate is sustained even at low substrate concentrations.
Analysis of a theoretical model consistent with the experimental results reveals that ATP synthesis implements a rotation-assisted mechanism, in which a limited binding rate at low substrate concentration is partially compensated for by rotation to an angle where the binding rate is high.
The results may imply that \Fone-ATPase implements a regulatory mechanism of enhancing substrate binding for ATP synthesis.

\end{abstract}

\maketitle

\section{Introduction}

Enzymes are catalysts that maintain biological activity by regulating the rates of various reactions \textit{in vivo}.
Enzyme kinetics, that is, the dependence of reaction rate \(V\) on substrate concentration \([\Subs]\), is essential for characterizing enzymes \cite{BD02062988}.
The Michaelis-Menten equation,
\begin{align}
 V = \frac{\Vmax [\Subs]}{\Km + [\Subs]},
 \label{e:MM}
\end{align}
is a standard kinetics of enzymatic reactions, and the maximum rate \(\Vmax\) and Michaelis constant \(\Km\) are often used to characterize these reactions.
Deviations from Eq.~(\ref{e:MM}) imply the existence of regulatory mechanisms such as the cooperativity between substrate binding sites.
Enzyme kinetics have been used to investigate the reaction mechanisms of a molecular motor \Fone-ATPase (\Fone).

\Fone is an enzyme that synthesizes \ac{ATP}, which is often referred to as the currency of energy in cells, in aerobic respiration and photosynthesis and plays an essential role in energy metabolism.
\Fone has three catalytic sites located at three interfaces between \(\alpha\) and \(\beta\) subunits, which consist of the \(\alpha_3\beta_3\) ring surrounding the \(\gamma\) subunit \cite{abrahams_structure_1994}.
Another motor F\textsubscript{o} drives \Fone by applying an external torque, and \Fone synthesizes ATP \textit{in vivo},
whereas \Fone catalyzes ATP hydrolysis when the external torque is absent or small.
In this sense, \Fone is said to be reversible.

The enzyme kinetics of \Fone is perturbed by regulatory mechanisms.
\Fone employs multiple mechanisms to inhibit only ATP hydrolysis, including ADP inhibition, \(\epsilon\) subunit, and regulatory protein IF\textsubscript{1} \cite{lapashina_adp-inhibition_2018}, which pause the enzyme activity under conditions that promote ATP hydrolysis.
Since the primary role of \Fone in cells is ATP synthesis, it may be considered that these mechanism suppresses unfavorable futile ATP consumption by ATP hydrolysis. 
The deviation in the enzyme kinetics of \Fone \cite{ebel_stimulation_1975} was originally examined in the context of the cooperativity of three catalytic sites \cite{schuster_kinetic_1975,chernyak_structural_1981,cross_mechanism_1982,gresser_catalytic_1982,wong_kinetics_1984,roveri_steadystate_1985}, and was later revealed to be caused by the regulatory effect of non-catalytic sites \cite{jault_slow_1993}.

Here, we show by single-molecule experiments that the enzyme kinetics of \Fone is itself asymmetric even without the pause of the enzyme activity discussed above.
We see that this asymmetric kinetics originates from a rotation-assisted substrate binding in ATP synthesis.
This novel mechanism is derived based on a previously proposed model with asymmetric allosterism \cite{zimmermann_efficiencies_2012,kawaguchi_nonequilibrium_2014}.

\section{Materials and Methods}
\subsection{Preparation of \Fone}

$\alpha_3\beta_3\gamma$ subcomplexes of \Fone derived from a thermophilic \textit{Bacillus} PS3 with mutations for the rotation assay (His$_6$-$\alpha$C193S/W463F, His$_{10}$-$\beta$, and $\gamma$S107C/I210C) \cite{rondelez_highly_2005} were expressed and purified in the same way as \cite{nakayama_optimal_2021}, except that the treatment with (\(\pm\))-dithiothreitol before flash freezing was omitted.

\subsection{Single-molecule experiments}
We observed the rotation of \Fone in single-molecule experiments under an external torque \(\Nex\) by using an electrorotation method \cite{washizu_dielectrophoretic_1993,watanabe-nakayama_effect_2008} instead of F\textsubscript{o} motor.
The experimental setup is basically the same as that in \cite{nakayama_optimal_2021}.
The \Fone molecules were fixed on a Ni-NTA-modified coverslip, which is part of an observation chamber.
The rotation of a single \Fone molecule was visualized by attaching a probe particle to its rotary shaft (\(\gamma\) subunit).
The rotation was observed with \SI{5}{mM} MOPS and \SI{1}{mM} phosphate buffer containing \SI{1}{mM} magnesium chloride, and indicated amount of MgATP and MgADP (pH 7.0) at room temperature (\(25\pm 1\si{\degreeCelsius}\)).
See Appendix \ref{ss:setup} for details.

We applied an external torque in the clockwise direction when viewed from the protruded rotary shaft of \Fone.
\Fone rotated its rotary shaft clockwise under a sufficiently strong torque \cite{toyabe_thermodynamic_2011}, whereas the shaft was rotated counterclockwise in the absence of torque \cite{noji_direct_1997}.
In the clockwise rotation, it was confirmed that \Fone synthesizes ATP \cite{itoh_mechanically_2004,rondelez_highly_2005}.
The rotation and the external torque are defined to be positive in the counterclockwise direction throughout this paper.

In order to examine the enzyme kinetics, the reaction rate should be measured as a function of the substrate concentration.
However, the substrates of ATP synthesis, \ac{ADP} and inorganic phosphate (\iP), differ from that of ATP hydrolysis, \ac{ATP} (and water).
Then, we chose to change \([\ATP]\) and \([\ADP]\) while keeping \([\ATP] = [\ADP]\) and \([\iP] = \SI{1}{mM}\).
We denote \([\Subs]:= [\ATP] = [\ADP]\).
This choice keeps the driving force of ATP hydrolysis constant, which is given by the free energy change associated with the hydrolysis of a single ATP molecule
\begin{align}
 \Delta\mu = \Delta\mu^\circ + \kb T \ln \frac{[\ATP]}{[\ADP][\iP]},
\end{align}
except for a slight variation in \(\Delta\mu^\circ\).
Here, \(\Delta\mu^\circ\) is the standard free energy change, \(\kb\) is the Boltzmann constant, and \(T\) is the ambient temperature.
The values of \(\Delta\mu\) were calculated by a method described in \cite{krab_improved_1992} as \(18.2 \kb T = \SI{75}{pN.nm}\) under all the conditions we examined.

\subsection{Model}
\subsubsection{Totally asymmetric allosteric model (TASAM)} \label{sss:TASAM}
The motion of molecular motors is often described by Brownian motion in stochastically switching potentials \cite{elston_energy_1998}.
In such a potential switching model of \Fone, the shaft angle \(\theta\) follows the overdamped Langevin equation
\begin{align}
 \Gamma \dot \theta = -\pdv{U_n(\theta)}{\theta} + \Nex + \sqrt{2\Gamma \kb T} \xi
 ,
\end{align}
where \(\Gamma\) is the rotational frictional coefficient, \(U_n(\theta)\) are potentials labeled by a discrete variable \(n\), and \(\xi\) is white Gaussian noise with zero mean and unit variance.
The change in the discrete variable \(n\) corresponds to the elementary reactions of \Fone, such as ATP (ADP) binding followed by ADP (ATP) release, binding and release of \iP, and the synthesis and hydrolysis of ATP at the binding sites.
Here, we considered only ATP binding, followed by ADP release and its reverse reaction, as the others occur rapidly under our experimental conditions \cite{yasuda_resolution_2001}.
These binding events increase and decrease \(n\) by one with rates \(R_n^+(\theta)\) and \(R_n^-(\theta)\), respectively.
The \ac{TASAM} \cite{kawaguchi_nonequilibrium_2014} is a potential switching model where \(R_n^+(\theta)\) and \(R_n^-(\theta)\) are set as
\begin{align}
 \begin{split}
  R_n^+(\theta) &= W
  ,\\
  R_{n+1}^-(\theta) &= W \exp\left[-\frac{U_{n}(\theta) - U_{n+1}(\theta) + \Delta\mu}{\kb T}\right]
  ,
 \end{split}
 \label{e:W of TASAM}
\end{align}
where \(W\) is assumed to be proportional to \([\Subs]\) as \(W = \lambda [\Subs]\), as it characterizes the rates of these binding events.
The angular dependence of \(R_n^\pm(\theta)\) indicates the allosterism between the rotary shaft and the binding sites and represents how much the angle \(\theta\) regulates the binding rates.
The angular dependence of the ratio, \(R_n^+(\theta) / R_{n+1}^-(\theta)\), imposed by local detailed balance, is entirely assigned to \(R_{n+1}^-(\theta)\).
In this sense, this model is referred to as totally asymmetric allosteric.
It was shown in Ref.~\cite{kawaguchi_nonequilibrium_2014} that the \ac{TASAM} quantitatively reproduces the experimental result of \Fone \cite{toyabe_nonequilibrium_2010}
(a similar argument was made in \cite{zimmermann_efficiencies_2012}, where the angular dependence of \(R_n^\pm(\theta)\) originates from a load-sharing factor).
We chose \(U_0(\theta)\) as
\begin{align}
 U_0(\theta) =& - \kb T \ln \left[
 \exp\left(- \frac{k \theta^2}{2\kb T}\right)
 \right.
 \nonumber \\
 &\left.+
 \exp\left(- \frac{k (\theta + \SI{40}{\degree})^2}{2\kb T} - \tdm\right)
 \right]
 ,
 \label{e:effective potential}
\end{align}
which is an effective potential obtained by coarse-graining two harmonic potentials with a common spring constant \(k\), angular displacement \SI{40}{\degree}, and height difference \(\tdm\) \cite{kawaguchi_nonequilibrium_2014}.
\(U_n(\theta)\) is obtained from \(U_0(\theta)\) by imposing a translational symmetry \(U_n(\theta) = U_{n+1}(\theta+\SI{120}{\degree})\), which corresponds to
the threefold symmetry of the \(\alpha_3\beta_3\) subcomplex surrounding the rotary shaft of \Fone.

\subsubsection{Fine-grained model} \label{sss:fine-grained}
It is necessary to take into account the other elementary reactions than ATP (ADP) binding followed by ADP (ATP) release, when they are not rapid.
For this purpose, we here split each state labeled by \(n\) into two states corresponding to the ATP hydrolysis dwell (\(i = \h\)) and the ATP binding dwell (\(i = \B\)).
The potentials are labeled by \(n\) and \(i\), and given as
\begin{align}
    U_{0,\B}(\theta) &= \frac{k}{2} \theta^2, 
    &
    U_{0,\h}(\theta) &= \frac{k}{2} (\theta + \SI{40}{\degree})^2 + \tdm,
\end{align}
with a translational symmetry \(U_{n,i}(\theta) = U_{n+1,i}(\theta+\SI{120}{\degree})\).
The time evolution of \(\theta\) follows the overdamped Langevin equation
\begin{align}
    \Gamma \dot \theta = -\pdv{U_{n,i}(\theta)}{\theta} + \Nex + \sqrt{2\Gamma \kb T} \xi.
\end{align}
The rates of switching, from \((n,\B)\) to \((n+1,\h)\), from \((n+1,\h)\) to \((n,\B)\), from \((n,\h)\) to \((n,\B)\), and from \((n,\B)\) to \((n,\h)\)
were set as
\begin{align}
    \begin{split}
    &R^+_{n,\B}(\theta) = W,
    \\
    &R^-_{n+1,\h}(\theta) = W \exp\left[-\frac{U_{n,\B}(\theta) - U_{n+1,\h}(\theta) + \Delta\mu}{\kb T}\right],
    \\
    &R^+_{n,\h}(\theta) = W' \exp\left[q'\frac{U_{n,\h}(\theta) - U_{n,\B}(\theta)}{\kb T}\right],
    \\
    &R^-_{n,\B}(\theta) = W' \exp\left[-(1-q')\frac{U_{n,\h}(\theta) - U_{n,\B}(\theta)}{\kb T}\right],
    \end{split}
\end{align}
respectively.
Here, \(W\) is proportional to \([\Subs]\), since \(R^+_{n,\B}\) and \(R^-_{n+1,\h}\) are the rates of ATP binding followed by ADP release and its reverse reaction, respectively.
On the other hand, \(W'\) is independent on \([\Subs]\), since \(R^+_{n,\h}\) and \(R^-_{n,\B}\) are the rates of other elementary reactions that do not involve the binding of either ATP or ADP.
Although two elementary reactions occur sequentially at the switching between \((n,\h)\) and \((n,\B)\) \cite{yasuda_resolution_2001}, we regard them as a single reaction for simplicity.
\(q'\) is a parameter to control the angular dependence of \(R^+_{n,h}(\theta)\) and \(R^-_{n,b}(\theta)\).
The ratios, \(R^+_{n,\B}(\theta) / R^-_{n+1,\h}(\theta)\) and \(R^+_{n,\h}(\theta) / R^-_{n,\B}(\theta)\), satisfy local detailed balance condition.
This model is coarse-grained to the TASAM in the limit of \(W'\to\infty\) irrespective of the value of \(q'\) \cite{kawaguchi_nonequilibrium_2014}.

\section{Results}
\subsection{Enzyme kinetics for ATP synthesis and hydrolysis}
\begin{figure}
 \includegraphics{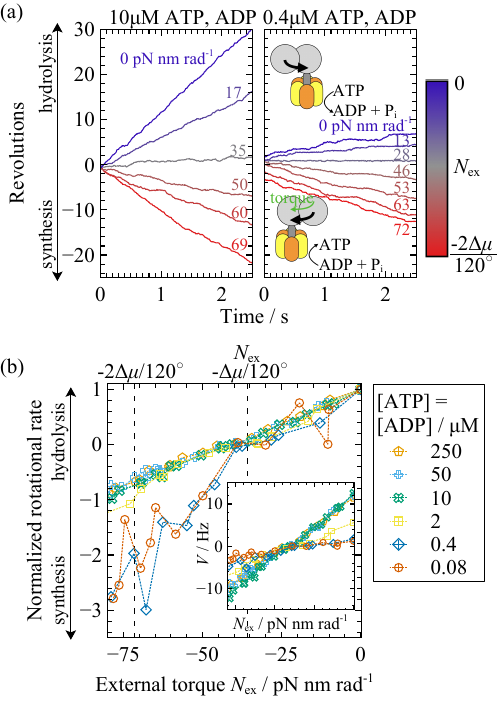}
 \caption{
 a) Single-molecule trajectories in the absence and presence of external torque. The values of \([\ATP] = [\ADP]\) are as indicated. \([\iP] = \SI{1}{mM}\).
 The values of the external torque are indicated on the sides of the trajectories and by the line colors.
 b) Examples of dependence of \(V\) on \(\Nex\). \(V\) are normalized by their values at \(\Nex = 0\). The values of \([\ATP] = [\ADP]\) are as indicated. \([\iP] = \SI{1}{mM}\). Data were obtained for one molecule at each concentration of ATP and ADP.
 Inset: Dependence of \(V\) on \(\Nex\) without the normalization of \(V\). 
 }
 \label{f:el}
\end{figure}
The dependence of the rotational rate \(V\) on external torque \(\Nex\) differs depending on \([\Subs]\) as shown in Fig.~\ref{f:el}.
For example, we compare the two cases: \(\Nex = 0\) and \(\Nex\simeq-2\Delta\mu/\SI{120}{\degree}\) (\SI{-72}{pN nm/rad} for \([\Subs]=\SI{0.4}{\micro M}\) and \SI{-76}{pN nm/rad} for \SI{10}{\micro M}).
The net driving force per \SI{120}{\degree} is given as \(\Nex \times \SI{120}{\degree} + \Delta\mu\), since the \SI{120}{\degree} rotation of \Fone is tightly coupled with the synthesis or hydrolysis of one ATP molecule \cite{yasuda_f1-atpase_1998,nishizaka_chemomechanical_2004,rondelez_highly_2005,toyabe_thermodynamic_2011,toyabe_single_2015,saita_simple_2015,soga_perfect_2017}.
Hence, the two cases are chosen so that the net driving force has the same absolute value but the opposite sign, \(\pm\Delta\mu\).
We typically observe that \(|V|\) for \(\Nex = 0\) is smaller than that for \(\Nex\simeq-2\Delta\mu/\SI{120}{\degree}\) with \([\Subs] = \SI{0.4}{\micro M}\) [Fig.~\ref{f:el}(a), right panel],
but \(|V|\) for \(\Nex = 0\) is larger than that for \(\Nex\simeq-2\Delta\mu/\SI{120}{\degree}\) with \([\Subs] = \SI{10}{\micro M}\) [Fig.~\ref{f:el}(a), left panel].
The rotational rates normalized by the rates with \(\Nex=0\) show significant change in the dependence of \(V\) on \(\Nex\) depending on \([\Subs]\) [Fig.~\ref{f:el}(b)].
The normalized rotational rates exhibit asymmetric torque dependence between ATP synthesis and hydrolysis for \([\Subs]\) less than approximately \SI{2}{\micro M} [Fig.~\ref{f:el}(b)], while they collapse into a single symmetric curve for larger \([\Subs]\). 
Long pauses caused by ADP inhibition \cite{noji_direct_1997, hirono-hara_pause_2001, hirono-hara_activation_2005,nakayama_optimal_2021} were excluded from the calculation of \(V\) to focus on the bare enzyme kinetics of \Fone.
Therefore, these results suggest that \Fone's enzyme kinetics is itself asymmetric between ATP synthesis and hydrolysis.

\begin{figure}
 \includegraphics{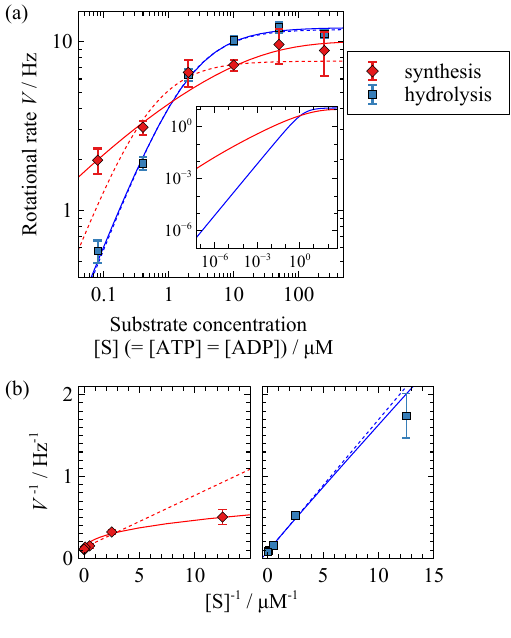}
 \caption{Dependence of the rotational rate on substrate concentration in ATP synthesis and hydrolysis.
 a) Experimental and numerical results.
 The symbols represent the experimental results.
 Error bars denote standard error of the mean.
 Each data point is the average of 8-9 molecules.
 The dashed curves are fitted to the experimental results using Eq.~(\ref{e:MM}).
 The solid curves are the numerical results obtained from the TASAM.
 The values of the parameters are
 \(k = 15.4 \kb T / \si{rad^2}\), \(\tdm = 1.45 \kb T\), \(\lambda = 24.4 / \si{\micro M.s}\), and \(\Gamma = \SI{0.29}{pN.nm.s/rad^2}\).
 The sum of weighted squared residuals for the numerical results is about half that for the fitting with Eq.~(\ref{e:MM}).
 Inset: Numerical results for a wide range of \([\Subs]\), part of which is experimentally inaccessible.
 b) Lineweaver-Burk plot.
 The symbols represent the experimental results.
 Error bars denote standard error of the mean.
 The dashed lines represent \(V^{-1} = \Vmax^{-1} (\Km [\Subs]^{-1} + 1)\).
 The solid curves correspond to the numerical results shown in a).
 }
 \label{f:experimental}
\end{figure}

We measured the enzyme kinetics with \(\Nex = -2\Delta\mu/\SI{120}{\degree}\) for ATP synthesis and \(\Nex = 0\) for ATP hydrolysis [Fig.~\ref{f:experimental}].
Because it is difficult to precisely tune the value of \(\Nex\), \(V\) is calculated by interpolating the rotation rates in the vicinity of \(\Nex = -2\Delta\mu/\SI{120}{\degree}\) for ATP synthesis.
We find that ATP hydrolysis follows Michaelis-Menten equation.
Equation~(\ref{e:MM}) fits \(V\) in ATP hydrolysis with the fitting parameters \(\Vmax = \SI{11.8}{Hz}\) and \(\Km = \SI{1.9}{\micro M}\).
This result is qualitatively the same as that in the absence of ADP \cite{
yasuda_f1-atpase_1998,
yasuda_resolution_2001,
noji_purine_2001,
bandyopadhyay_g156c_2002,
sakaki_one_2005,
adachi_coupling_2007,
watanabe_temperature-sensitive_2008,
furuike_temperature_2008,
tsumuraya_effect_2009,
okuno_single-molecule_2013,
yukawa_key_2015,
watanabe_essential_2018%
}. 
In ATP synthesis, \(V\) also decreases as \([\Subs]\) decreases; however the dependence of \(V\) on \([\Subs]\) is weaker than that in ATP hydrolysis.
Equation~(\ref{e:MM}) cannot describe this characteristic because its curve on a log-log plot exhibits such a weak dependence on \([\Subs]\) only within a narrow range of \([\Subs]\) around \(\Km\).
The changes in \(\Vmax\) and \(\Km\) simply translate the curve while keeping its shape.
The deviation from the Michaelis-Menten equation (\ref{e:MM}) can also be observed in the Lineweaver-Burk plot [Fig.~\ref{f:experimental}(b)], where a linear plot is expected for the Michaelis-Menten equation.
We observed a convex plot for the ATP synthesis and a linear plot for the ATP hydrolysis.
These results indicate the asymmetric enzyme kinetics of \Fone between ATP synthesis and hydrolysis in the sense that their functional forms are different.

\begin{figure}
\includegraphics{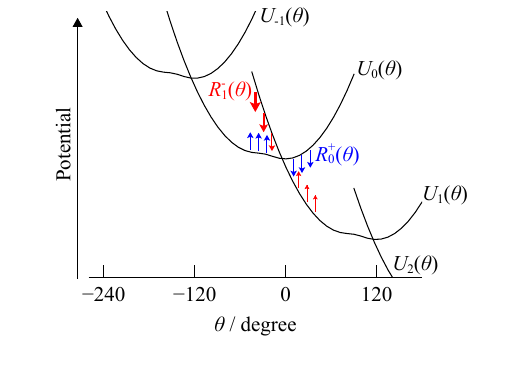}
\caption{Schematic of the TASAM.
Solid black curves represent \(U_n(\theta)\).
\(U_n(\theta)\) are plotted with vertical shifts \(n\Delta\mu\).
\(R_0^+(\theta)\) and \(R_1^-(\theta)\) are shown as blue and red arrows, respectively.
The thicknesses of the arrows represent the magnitude of \(R_0^+(\theta)\) and \(R_1^-(\theta)\).
}
\label{f:TASAM}
\end{figure}

Next, we investigated the asymmetry of the enzyme kinetics between ATP synthesis and hydrolysis based on TASAM (See \S\ref{sss:TASAM} for details).
A schematic of the \ac{TASAM} is shown in Fig.~\ref{f:TASAM}.
We numerically solved a master equation of TASAM to obtain the steady state distribution of \(\theta\) and \(n\) and evaluated \(V\) as a function of \([\Subs]\) (See Appendix \ref{ss:a_TASAM} for details).
We adjusted the values of \(k\), \(\tdm\), \(\lambda\), and \(\Gamma\) to minimize the sum of weighted squared residuals between the numerical results of \(V\) and the experimental results of \(V\).
Here, the weights are set to the inverse square of the standard error of the mean.
As can be seen in Fig.~\ref{f:experimental}, the numerical results showed better agreement with the experimental results than the fitting with Eq.~(\ref{e:MM}).
In addition, the numerical results reproduced two characteristics of the experimental results: the weak dependence of \(V\) on \([\Subs]\) in ATP synthesis, which Eq.~(\ref{e:MM}) cannot describe, and the enzyme kinetics following Eq.~(\ref{e:MM}) in ATP hydrolysis [Fig.~\ref{f:experimental}(a)].
Therefore, the theoretical model is consistent with the experimental results that the enzyme kinetics of \Fone deviate from Eq.~(\ref{e:MM}) in ATP synthesis, whereas it follows Eq.~(\ref{e:MM}) in ATP hydrolysis.
The present results extend the applicability of the TASAM, which was established mainly in the ATP hydrolysis \cite{kawaguchi_nonequilibrium_2014}, to the ATP synthesis.

\subsection{Relation with switching angle distribution}
\begin{figure}
 \includegraphics{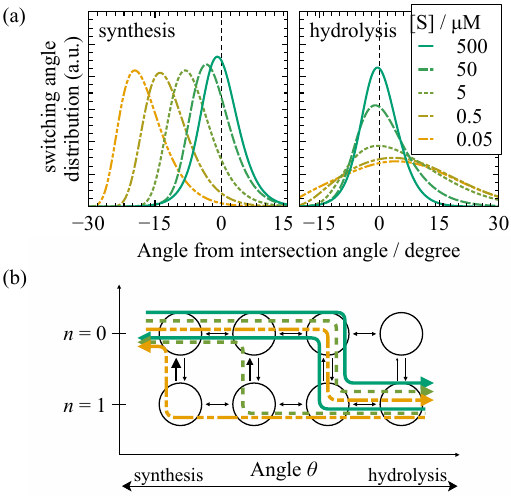}
 \caption{Switching angle distribution with a schematic of the path in a space of \(\theta\) and \(n\) in ATP synthesis and hydrolysis.
 The switching angle distribution is given as a difference between the angular distribution of the forward switching from \(n=0\) to \(n=1\) and that of the backward switching from \(n=1\) to \(n=0\).
 The heights of the switching angle distributions are normalized.
 The origin of the horizontal axis is set at an intersection point of potentials. 
 The parameters in the numerical simulations are the same as those shown in Fig.~\ref{f:experimental}.
 In the schematic, the angle is represented as a discrete variable.
 The thicknesses of the vertical arrows represent the magnitude of \(R_0^+(\theta)\) and \(R_1^-(\theta)\).
 }
 \label{f:switching angle}
\end{figure}

We focus on the switching angle distribution to elucidate the mechanism underlying the asymmetry in the enzyme kinetics by using the TASAM [Fig.~\ref{f:switching angle}].
It was shown that the dependence of the switching angle distribution on \([\Subs]\) is asymmetric between ATP synthesis and hydrolysis \cite{kawaguchi_nonequilibrium_2014}.
We numerically calculated the switching angle distribution at several concentrations \([\Subs]\) using the steady state distribution of \(\theta\) and \(n\) (See Appendix \ref{ss:a_TASAM} for details).
The peak position of the switching angle distribution is kept almost constant around the intersection point of potentials, where \(U_{n}(\theta) - U_{n+1}(\theta) + \Delta\mu = 0\),
irrespective of \([\Subs]\) in ATP hydrolysis,
but shifts in the negative direction in ATP synthesis as \([\Subs]\) decreases as discussed in Ref.~\cite{kawaguchi_nonequilibrium_2014}.
The switching around the intersection point in ATP hydrolysis was experimentally observed
\cite{toyabe_recovery_2012}.
The peak shift in ATP synthesis
suggests that a path in a space of \(\theta\) and \(n\) changes with \([\Subs]\) (bottom panel in Fig.~\ref{f:switching angle}).
At high \([\Subs]\), ADP binds quickly around the intersection point, and then the shaft rotates.
In contrast, at low \([\Subs]\), the shaft rotates beyond the intersection point, and then ADP binds.
That is, the path where substrate binding is slow is bypassed by the uphill rotation of \Fone.
This bypass does not occur in ATP hydrolysis, because the binding rate of ATP, \(R_n^+(\theta)\), does not increase, even if \Fone climbs up the potential slope [Eq.~(\ref{e:W of TASAM})].

This difference in the switching angle distributions between ATP synthesis and hydrolysis causes the asymmetry in the enzyme kinetics as follows.
In ATP hydrolysis, the peak position remains almost constant.
This means that the rates of ATP binding followed by ADP release and its reverse reaction are written as \(k_\ATP [\ATP]\) and \(k_\ADP [\ADP]\), respectively, with \(\theta\)-independent bimolecular rate constants,
\(k_\mathrm{ATP}\) and \(k_\mathrm{ADP}\).
Therefore, the reaction can be described using a sequential scheme
\begin{align}
\cdots
\xrightleftharpoons[k'_\mathrm{rot}]{k_\mathrm{rot}}
\bigcirc
\xrightleftharpoons[{k_\mathrm{ADP}[\ADP]}]{k_\mathrm{ATP}[\ATP]}
\bigcirc
\xrightleftharpoons[k'_\mathrm{rot}]{k_\mathrm{rot}}
\bigcirc
\cdots
,
\end{align}
and its enzyme kinetics follows Eq.~(\ref{e:MM}).
Here, \(k_\mathrm{rot}\) and \(k'_\mathrm{rot}\) are  rate constants corresponding to \SI{\pm120}{\degree} rotation.
In contrast, the uphill rotation in ATP synthesis assisted by thermal diffusion partially compensates for the decrease in the substrate binding rate.
As a result, the dependence of \(V\) on \([\Subs]\) in ATP synthesis is weaker than that in ATP hydrolysis.
Since \(V\) is proportional to \([\Subs]\) when the binding of the substrate is rate-limiting, this result indicates that \Fone prevents the binding of ADP from becoming a rate-limiting step.
That is, the asymmetric allosterism represented by Eq.~(\ref{e:W of TASAM}) causes the difference in the switching angle distributions, which results in the asymmetry in the enzyme kinetics.

\subsection{asymmetric allosterism and asymmetric potential}
\begin{figure}
 \includegraphics{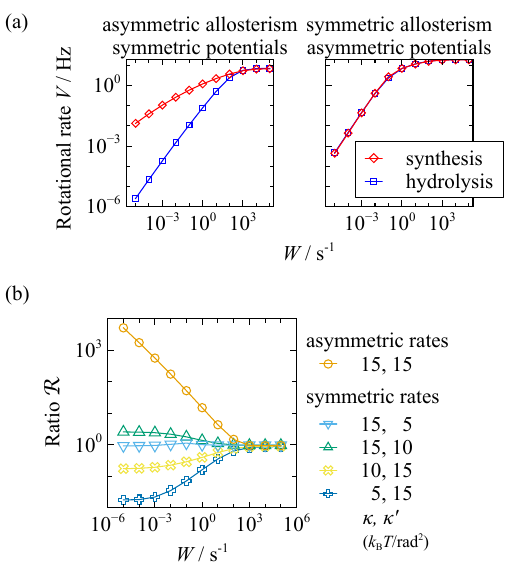}
 \caption{Comparison between asymmetric allosterism with symmetric potentials, and symmetric allosterism with asymmetric potentials.
 a) Dependence of \(V\) on \(W\) in ATP synthesis and hydrolysis. \(\kappa = \kappa' = 15 \kb T / \si{rad^2}\) for symmetric potentials, and \(\kappa = 15 \kb T / \si{rad^2}\), \(\kappa' = 5 \kb T / \si{rad^2}\) for asymmetric potentials.
 b) Dependence of ratios of \(V\) in ATP synthesis to \(V\) in ATP hydrolysis, \(\mathcal{R}\), on \(W\).
 }
 \label{f:origin of asymmetry}
\end{figure}
We explore other possibilities of generating asymmetric kinetics than the asymmetric allosterism.
The asymmetric kinetics provide a simple regulatory mechanism for enzymatic reactions and may be used by other molecular motors. It may also be important in applications of developing novel enzymes.
In particular, we compare two cases: asymmetric allosterism with symmetric potentials and symmetric allosterism with asymmetric potentials.
Symmetric rates
\begin{align}
 \begin{split}
  R_n^+(\theta) &= W \exp\left[\frac{U_{n}(\theta) - U_{n+1}(\theta) + \Delta\mu}{2\kb T}\right]
  ,\\
  R_{n+1}^-(\theta) &= W \exp\left[-\frac{U_{n}(\theta) - U_{n+1}(\theta) + \Delta\mu}{2\kb T}\right]
 \end{split}
 \label{e:symmetric W}
\end{align}
were used for the symmetric allosterism.
Asymmetric potentials were constructed by concatenating harmonic potentials with different spring constants \(\kappa\) and \(\kappa'\),
\begin{align}
 U_0(\theta) = 
 \begin{cases}
  \frac{1}{2} \kappa \theta^2 & \theta \geq 0,
  \\
  \frac{1}{2} \kappa' \theta^2 & \theta < 0,
 \end{cases}
\end{align}
and controlled its asymmetry through the ratio of \(\kappa\) and \(\kappa'\).
The enzyme kinetics for these cases were numerically evaluated [Fig.~\ref{f:origin of asymmetry}(a)].
The asymmetric allosterism with symmetric potentials induced asymmetric enzyme kinetics between ATP synthesis and hydrolysis, qualitatively reproducing those of the \ac{TASAM} [Fig.~\ref{f:experimental}(a)].
The enzyme kinetics for symmetric allosterism with asymmetric potentials were similar to the Michaelis-Menten kinetics [Eq.~(\ref{e:MM})] in that \(V\) was proportional to \(W\) at small \(W\).

The asymmetry of the enzyme kinetics can be measured by the ratio of \(V\) in ATP synthesis and hydrolysis, \(\mathcal{R}\) [Fig.~\ref{f:origin of asymmetry}(b)]. 
With asymmetric allosterism, \(\mathcal{R}\) continued to increase as \(W\) decreases, indicating the deviation of ATP synthesis from the Michaelis-Menten kinetics.
In contrast, with symmetric allosterism, \(\mathcal{R}\) converge as \(W\) decreases even with significant asymmetry in \(\kappa\) and \(\kappa'\), reflecting the fact that both ATP hydrolysis and synthesis approximately follow Michaelis-Menten kinetics.
Thus, the asymmetric allosterism is essential for the asymmetric enzyme kinetics. 

\subsection{Effect of other elementary reactions}
\begin{figure}
    \includegraphics{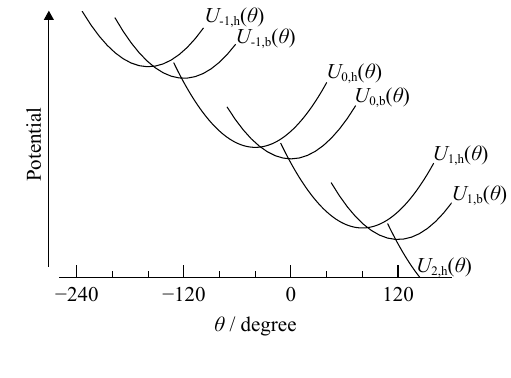}
    \caption{
    Schematic of the potentials of the fine-grained model. Solid black curves represent the potentials, \(U_{n,i}(\theta)\). \(U_{n,\B}(\theta)\) and \(U_{n,\h}(\theta)\) are plotted with vertical shifts \(n\Delta\mu\).
    }
    \label{f:fine-grained}
\end{figure}

Finally, we consider the physiological implications of the present work. 
In our experimental system, a relatively large probe, which is necessary for applying the external torque, is attached to the rotary shaft of \Fone.
\(\Gamma\) is thought to be significantly smaller in cells.
The above modeling by TASAM assumes that the binding and release of \iP and the synthesis and hydrolysis of ATP at the binding sites are sufficiently faster than the diffusion in the potentials and are not explicitly treated as independent mechanochemical steps.
This assumption is no longer effective for small \(\Gamma\) expected in physiological conditions, and more detailed modeling is required in such situations.
Therefore, we performed the numerical simulation of a fine-grained model that explicitly takes into account these elementary reactions (See \S\ref{sss:fine-grained} and Appendix \ref{ss:a_fine-grained} for details).
A schematic of the potentials of this model is shown in Fig.~\ref{f:fine-grained}.

\begin{figure*}
    \includegraphics{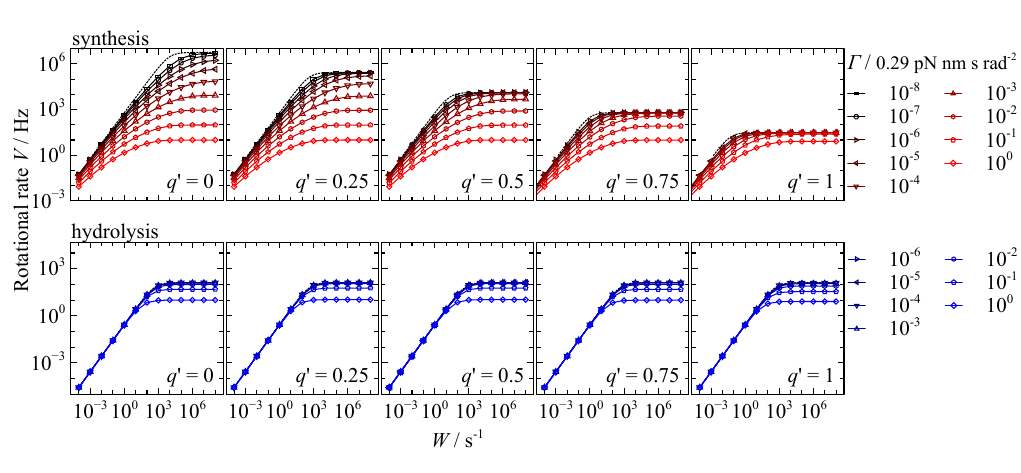}
    \caption{
    Dependence of enzyme kinetics on \(\Gamma\) in ATP synthesis and hydrolysis obtained from the fine-grained model.
    The values of \(q'\) are as indicated.
    The value of \(W'\) for each \(q'\) is set so that the rotational rates in the limit of small \(\Gamma\) and high \([\Subs]\) equal to the experimental value, \SI{129}{Hz} \cite{yasuda_resolution_2001}.
    The dashed curves represent the Michaelis-Menten kinetics obtained in the limit of \(\Gamma\to0\) (See Appendix \ref{sss:low frictional limit} for details).
    As Fig.\ref{f:experimental}, \(k = 15 \kb T / \si{rad^2}\) and \(\tdm = 2.6 \kb T\).
    }
    \label{f:crossover}
\end{figure*}
The enzyme kinetics in ATP synthesis converges to the Michaelis-Menten kinetics as \(\Gamma\) decreases [Fig.~\ref{f:crossover}].
Since there is no experimental knowledge on \(q'\), we here varied \(q'\) between 0 and 1.
The convergence to the Michaelis-Menten kinetics is observed irrespective of the value of \(q'\).
The convergence indicates that the angular distribution approaches equilibrium in each potential.
This result also supports the rotation-assisted mechanism; the decrease in the substrate binding rate is not compensated when diffusion is much faster than the substrate binding, since the uphill rotation takes place irrespective of \([\Subs]\).
This convergence to the Michaelis-Menten kinetics is consistent with previous experiments reporting that ATP synthesis by \FoFone reconstituted into liposomes follows the Michaelis-Menten kinetics \cite{bald_atp_1998,iino_mechanism_2009,soga_efficient_2011}.
Therefore, the enzyme kinetics observed in our experiment continuously changes to Michaelis-Menten kinetics by decreasing \(\Gamma\).
We emphasize that the TASAM does not exhibit such behavior, because the change in the value of \(\Gamma\) does not affect the dependence of \(\Gamma V\) on \(\Gamma [\Subs]\).
That is, the elementary reactions such as the binding and release of \iP, and the synthesis and hydrolysis of ATP are essential to explain the dependence of the enzyme kinetics on \(\Gamma\).

Temperature is another important factor different from physiological one, since \Fone we used was derived from thermophile whose optimal temperature is around \SI{60}{\degreeCelsius}.
The optimal temperature is significantly higher than the temperatures in our experiments (\SI{25}{\degreeCelsius}) and the previous work (24--\SI{40}{\degreeCelsius}) \cite{bald_atp_1998,iino_mechanism_2009,soga_efficient_2011}.
Since the elementary reactions are expected to become faster at higher temperatures, the assumption of the TASAM may be recovered, and the enzyme kinetics may deviate from the Michaelis-Menten equation (\ref{e:MM}) even for small \(\Gamma\) at the physiological temperature.
However, the ATP-synthetic experiments for small \(\Gamma\) and at high temperatures are quite challenging and remain for future studies.

\section{Conclusion}
In this paper, we reveal that the enzyme kinetics of \Fone are asymmetric between ATP synthesis and hydrolysis by single-molecule experiments.
Here, we demonstrate that the model established mainly in the ATP hydrolysis \cite{kawaguchi_nonequilibrium_2014} is also applicable to the ATP-synthetic rotations and leads to asymmetric enzyme kinetics by a rotation-assisted mechanism where the rotational diffusion partially compensates the decrease in the substrate binding rate.
Our findings illustrate how the mechanochemical coupling implements a regulatory mechanism for enzymatic reactions.

\section{Acknowledgements}
This work was supported by JSPS KAKENHI Grant Numbers JP18H05427, JP19H01864, and JP23H01136, and JST ERATO Grant Number JPMJER2302.

\appendix

\section{Details of experimental setup} \label{ss:setup}
\subsection{Ni\(^{2+}\)-NTA modification of coverslip}
In our single-molecule experiments, \Fone molecules were adhered to a coverslip modified by Ni\(^{2+}\)-NTA.
The Ni\(^{2+}\)-NTA coverslip was prepared as follows.
Coverslips (\(24\times36\si{mm^2}\) or \(24\times32\si{mm^2}\), thickness No.~1; Matsunami) were washed with ultrapure water three times and immersed in \SI{10}{M} potassium hydroxide solution overnight.
The coverslips were washed five times with reverse osmosis (RO) water and once with ultrapure water, and immersed in a 1:1 mixture of ethanol and ultrapure water containing \SI{2}{\percent} v/v (3-Mercaptopropyl)trimethoxysilane (purchased from TCI, Japan) and \SI{0.06}{\percent} v/v acetic acid for \SI{120}{min} at \SI{60}{\degreeCelsius}.
The coverslips were washed three times with ultrapure water, and baked at \SI{120}{\degreeCelsius} for \SI{60}{min}.
Then, the coverslips were immersed in \SI{50}{mM} phosphate buffer (pH 7) containing \SI{1}{mM} (\(\pm\))-dithiothreitol and \SI{2}{mM} ethylenediaminetetraacetic acid for \SI{90}{min}.
Then, the coverslips were washed five times with RO water and once with ultrapure water, and reacted with \SI{1.8}{mg/ml} Maleimido-C\textsubscript{3}-NTA (purchased from Dojindo, Japan) in \SI{100}{mM} phosphate buffer (pH 7) for \SI{60}{min}.
Finally, the coverslips were immersed in \SI{50}{mM} nickel sulfate solution for \SI{30}{min} and washed five times with RO water and once with ultrapure water, before using the coverslip for single-molecule experiments.

\subsection{Single-molecule experiments}
The experimental setup is essentially the same as that in the previous study \cite{nakayama_optimal_2021}.
An observation chamber consisted of a Ni\(^{2+}\)-NTA modified coverslip and a slide with quadrupolar electrodes.
The coverslip and the slide were separated by double-sided adhesive tape (\SI{10}{\micro m} thickness; Teraoka, 7070W) with silicone grease (Shin-Etsu Chemical, Japan) on a surface.
The solution of \Fone was diluted with \SI{50}{mM} MOPS buffer (pH 7) containing \SI{5}{mg/ml} bovine serum albumin, \SI{50}{mM} potassium chloride, and \SI{1}{mM} magnesium chloride (buffer I) to a final concentration of \SI{1}{nM}.
The chamber was filled with the solution and incubated for \SI{10}{min} to immobilize \Fone molecules on the surfaces.
The bovine serum albumin was added as the blocking agent.
The chamber was washed with buffer I, and streptavidin-coated polystyrene particles (diameter = \SI{276}{nm}, Thermo Fisher Scientific) diluted with buffer I was infused into it.
Azide contained in the solution of polystyrene particle was removed in advance by repeating centrifugation, exchange of supernatant, and re-dispersion six times.
After a 30-min incubation, the solution in the chamber was exchanged with \SI{5}{mM} MOPS and \SI{1}{mM} phosphate buffer containing \SI{1}{mM} magnesium chloride, and indicated amount of MgATP and MgADP (pH 7.0).

Rotation of the $\gamma$-shaft was probed by dimeric polystyrene particles attached to the biotinylated $\gamma$-shaft.
The observation was performed on a bright-field upright microscope (Olympus, Japan) with a 100$\times$ objective (NA1.40), high-intensity LED (\SI{623}{nm}, \SI{4.8}{W}, Thorlabs, NJ) for illumination, a high-speed camera (Basler, Germany) at \(4,000\,\si{Hz}\), and a laboratory-made capturing software developed on LabVIEW (NI, TX).
The angular position of the dimeric probe was analyzed by an algorithm based on a principal component analysis of the probe image.
We applied an external torque on the probe attached to \Fone by using electrorotation method as described in~\cite{nakayama_optimal_2021}.
We kept the probe in focus by using a laboratory-made autofocus system that rotates a focus knob of the microscope by a stepping motor (Oriental motor, Japan) based on the real-time image analysis that quantifies the defocus of the probe image.

We observed 8, 9, 8, 8, 9, 9 molecules for \([\mathrm{ATP}] = [\mathrm{ADP}] = 0.08\), 0.4, 2, 10, 50, and \SI{250}{\micro M}, respectively.

\section{Details of numerical calculation}
\subsection{Numerical calculation of TASAM} \label{ss:a_TASAM}
We first obtained the probability distribution of \(\theta\) and \(n\) at the steady state \(P^\mathrm{st}(\theta, n)\). 
The master equation of the TASAM is written as
\begin{align}
 &\pdv{P_t(\theta, n)}{t}
 \nonumber \\
 &= -\pdv{}{\theta} \left[\frac{1}{\Gamma} \left( \Nex - \pdv{U_n(\theta)}{\theta}\right)  - \frac{\kb T}{\Gamma} \pdv{}{\theta}\right] P_t(\theta, n)
 \nonumber \\
 &\phantom{=}
 + R^+_{n-1}(\theta) P_t(\theta, n-1)
 + R^-_{n+1}(\theta) P_t(\theta, n+1)
 \nonumber \\
 &\phantom{=}
 - \left[R^+_n(\theta) + R^-_n(\theta) \right] P_t(\theta, n)
 ,
 \label{e:master}
\end{align}
where \(P_t(\theta, n)\) is the probability distribution of \(\theta\) and \(n\) at time \(t\).
\(P^\mathrm{st}(\theta, n)\) satisfies 
\begin{align}
 0 &= -\pdv{}{\theta} \left[\frac{1}{\Gamma} \left( \Nex - \pdv{U_0(\theta)}{\theta}\right)  - \frac{\kb T}{\Gamma} \pdv{}{\theta}\right] P^\mathrm{st}(\theta, 0)
 \nonumber \\
 &\phantom{=}
 + R^+_{0}(\theta+\SI{120}{\degree}) P^\mathrm{st}(\theta+\SI{120}{\degree}, 0)
 \nonumber \\
 &\phantom{=}
 + R^-_{0}(\theta-\SI{120}{\degree}) P^\mathrm{st}(\theta-\SI{120}{\degree}, 0)
 \nonumber \\
 &\phantom{=}
 - \left[R^+_0(\theta) + R^-_0(\theta) \right] P^\mathrm{st}(\theta, 0)
 ,
 \label{e:master-steady}
\end{align}
since \(P^\mathrm{st}(\theta, n)\) has a translational symmetry \(P^\mathrm{st}(\theta, n) = P^\mathrm{st}(\theta+\SI{120}{\degree}, n+1)\).
We introduced reflecting boundaries at \(\theta_b^+\) and \(\theta_b^-\), and discretized angular space with an interval of \(\SI{1}{\degree}\).
Then, we obtain \(P^\mathrm{st}(\theta, 0)\) by solving the discretized version of Eq.~(\ref{e:master-steady}) using Gaussian elimination.

The rotational rate \(V\) was evaluated from \(P^\mathrm{st}(\theta, 0)\) as follows.
\(V\) is proportional to a probability current in \(\theta\)-direction at the steady state 
\begin{align}
 J_\theta^\mathrm{st} = \sum_{n=-\infty}^\infty \left[\frac{1}{\Gamma} \left( \Nex - \pdv{U_n(\theta)}{\theta}\right)  - \frac{\kb T}{\Gamma} \pdv{}{\theta}\right]P^\mathrm{st}(\theta, n).
 \label{e:J}
\end{align}
More specifically, \(V = J_\theta^\mathrm{st} / 3\), when \(P^\mathrm{st}(\theta,0)\) is normalized so that \(\int \mathrm d\theta P^\mathrm{st}(\theta, 0) = 1\) holds.
Then, we evaluated \(J_\theta^\mathrm{st}\) by rewriting Eq.~(\ref{e:J}) as
\begin{align}
 J_\theta^\mathrm{st} = \sum_{n=-\infty}^\infty \left[\frac{1}{\Gamma} \left( \Nex - \pdv{U_n(\theta)}{\theta}\right) P^\mathrm{st}(\theta - n \cdot \SI{120}{\degree}, 0)
 \right.
 \nonumber \\
 \left.- \frac{\kb T}{\Gamma} \pdv{P^\mathrm{st}(\theta - n \cdot \SI{120}{\degree}, 0)}{\theta}\right],
\end{align}
and converted it to \(V\).

The switching angle distribution is defined as \(\Lambda_n(\theta) = R^+_n(\theta)P^\mathrm{st}(\theta, n) - R^-_{n+1}(\theta)P^\mathrm{st}(\theta, n+1)\).
We evaluated \(\Lambda_0(\theta)\) as
\begin{align}
 \Lambda_0(\theta) = R^+_0(\theta)P^\mathrm{st}(\theta, 0) - R^-_0(\theta - \SI{120}{\degree})P^\mathrm{st}(\theta - \SI{120}{\degree}, 0).
\end{align}
\(\Lambda_n(\theta)\) may be obtained as \(\Lambda_n(\theta) = \Lambda_0(\theta - n \cdot \SI{120}{\degree})\) from the translational symmetry.

\subsection{Numerical calculation of fine-grained model} \label{ss:a_fine-grained}

The master equation of the fine-grained model is written as
\begin{align}
 \begin{split}
 &\pdv{P_t(\theta, n, \B)}{t} 
 \\
 &= -\pdv{}{\theta} \left[\frac{1}{\Gamma} \left( \Nex - \pdv{U_{n,\B}(\theta)}{\theta}\right)  - \frac{\kb T}{\Gamma} \pdv{}{\theta}\right] P_t(\theta, n, \B)
 \\
 &\phantom{=}
 + R^+_{n,\h}(\theta) P_t(\theta, n,\h)
 + R^-_{n+1,\h}(\theta) P_t(\theta, n+1, \h)
 \\
 &\phantom{=}
 - \left[R^+_{n,\B}(\theta) + R^-_{n,\B}(\theta) \right] P_t(\theta, n, \B)
 ,
 \\
 &\pdv{P_t(\theta, n, \h)}{t} 
 \\
 &= -\pdv{}{\theta} \left[\frac{1}{\Gamma} \left( \Nex - \pdv{U_{n,\h}(\theta)}{\theta}\right) - \frac{\kb T}{\Gamma} \pdv{}{\theta}\right] P_t(\theta, n, \h)
 \\
 &\phantom{=}
 + R^+_{n-1,\B}(\theta) P_t(\theta, n-1,\B)
 + R^-_{n,\B}(\theta) P_t(\theta, n, \B)
 \\
 &\phantom{=}
 - \left[R^+_{n,\h}(\theta) + R^-_{n,\h}(\theta) \right] P_t(\theta, n, \h)
 ,
 \end{split}
 \label{e:fg-master}
\end{align}
where \(P_t(\theta, n, i)\) is the probability distribution of \(\theta\), \(n\), and \(i\) at time \(t\).
Similarly to the previous sub-section, the steady state distribution \(P^\mathrm{st}(\theta, n, i)\) satisfies
\begin{align}
 \begin{split}
 0 &= -\pdv{}{\theta} \left[\frac{1}{\Gamma} \left( \Nex - \pdv{U_{0,\B}(\theta)}{\theta}\right) - \frac{\kb T}{\Gamma} \pdv{}{\theta}\right] P^\mathrm{st}(\theta, 0, \B)
 \\
 &\phantom{=}
 + R^+_{0,\h}(\theta) P^\mathrm{st}(\theta, 0,\h)
 \\
 &\phantom{=}
 + R^-_{0,\h}(\theta - \SI{120}{\degree}) P^\mathrm{st}(\theta - \SI{120}{\degree}, 0, \h)
 \\
 &\phantom{=}
 - \left[R^+_{0,\B}(\theta) + R^-_{0,\B}(\theta) \right] P^\mathrm{st}(\theta, 0, \B)
 ,
 \\
 0 &= -\pdv{}{\theta} \left[\frac{1}{\Gamma} \left( \Nex - \pdv{U_{0,\h}(\theta)}{\theta}\right) - \frac{\kb T}{\Gamma} \pdv{}{\theta}\right] P^\mathrm{st}(\theta, 0, \h)
 \\
 &\phantom{=}
 + R^+_{0,\B}(\theta + \SI{120}{\degree}) P^\mathrm{st}(\theta + \SI{120}{\degree}, 0,\B)
 \\
 &\phantom{=}
 + R^-_{0,\B}(\theta) P^\mathrm{st}(\theta, 0, \B)
 \\
 &\phantom{=}
 - \left[R^+_{0,\h}(\theta) + R^-_{0,\h}(\theta) \right] P^\mathrm{st}(\theta, 0, \h)
 ,
 \end{split}
 \label{e:fg-master-steady}
\end{align}
since \(P^\mathrm{st}(\theta, n, i)\) has a translational symmetry \(P^\mathrm{st}(\theta, n, i) = P^\mathrm{st}(\theta+\SI{120}{\degree}, n+1, i)\).
We introduced reflecting boundaries, discretized angular space, and obtain \(P^\mathrm{st}(\theta, n, i)\) by solving the discretize version of Eq.~(\ref{e:fg-master-steady}).
In this model, a probability current in \(\theta\)-direction at the steady state is given as
\begin{align}
 J_\theta^\mathrm{st} = \sum_{n=-\infty}^\infty
 &\left[
 \frac{1}{\Gamma} \left( \Nex - \pdv{U_{n,\B}(\theta)}{\theta}\right) P^\mathrm{st}(\theta, n, \B)
 \right.
 \nonumber \\
 &- \frac{\kb T}{\Gamma} \pdv{P^\mathrm{st}(\theta, n, \B)}{\theta}
 \nonumber \\
 &+ \frac{1}{\Gamma} \left( \Nex - \pdv{U_{n,\h}(\theta)}{\theta}\right) P^\mathrm{st}(\theta, n, \h)
 \nonumber \\
 &\left.
 - \frac{\kb T}{\Gamma} \pdv{P^\mathrm{st}(\theta, n, \h)}{\theta}
 \right].
\end{align}
Then, we evaluated it as
\begin{align}
 J_\theta^\mathrm{st} = \sum_{n=-\infty}^\infty
 &\left[
 \frac{1}{\Gamma} \left( \Nex - \pdv{U_{n,\B}(\theta)}{\theta}\right) P^\mathrm{st}(\theta - n\cdot\SI{120}{\degree}, 0, \B)
 \right.
 \nonumber \\
 -& \frac{\kb T}{\Gamma} \pdv{P^\mathrm{st}(\theta - n\cdot\SI{120}{\degree}, 0, \B)}{\theta}
 \nonumber \\
 +&
 \frac{1}{\Gamma} \left( \Nex - \pdv{U_{n,\h}(\theta)}{\theta}\right) P^\mathrm{st}(\theta - n\cdot\SI{120}{\degree}, 0, \h)
 \nonumber \\
 -& \left.\frac{\kb T}{\Gamma} \pdv{P^\mathrm{st}(\theta - n\cdot\SI{120}{\degree}, 0, \h)}{\theta}
 \right],
\end{align}
and obtained \(V = J_\theta^\mathrm{st} / 3\).

\section{Enzyme kinetics of fine-grained model in the limit of \(\Gamma\to0\)} \label{sss:low frictional limit}
\begin{figure}[bp]
    \includegraphics[width=\hsize]{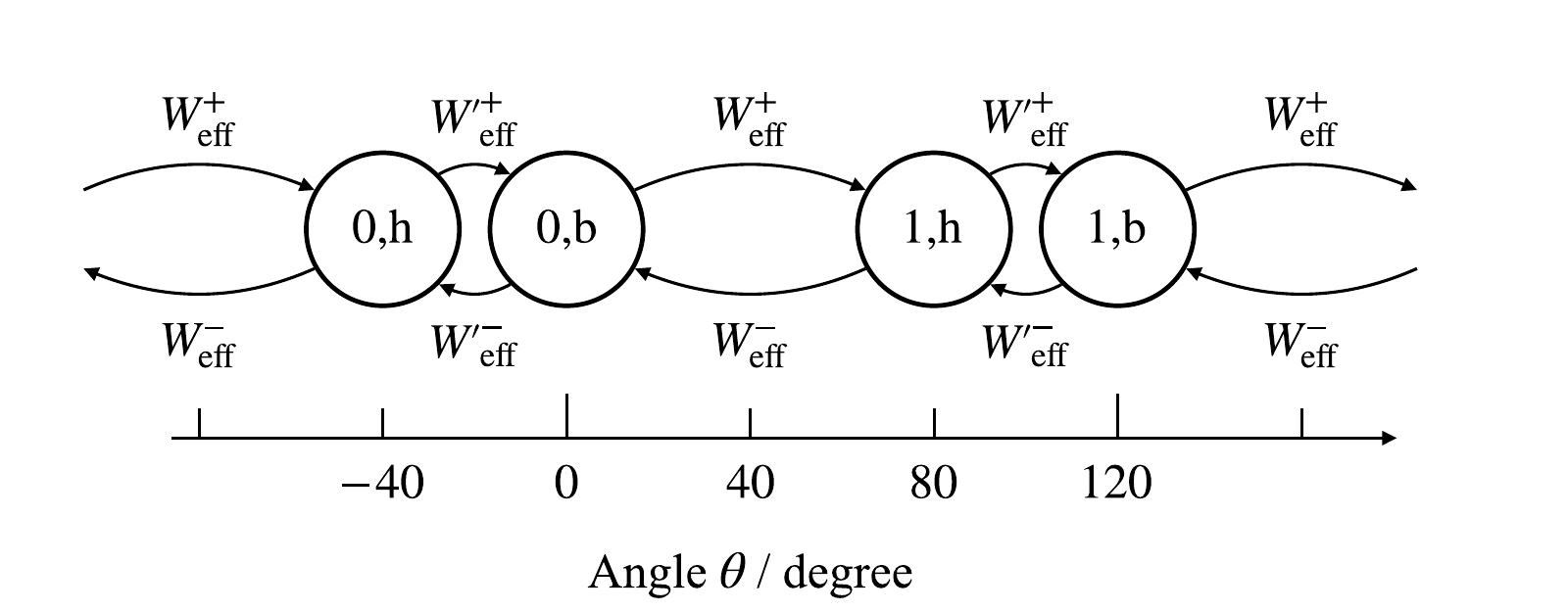}
    \caption{
    Schematic of Markov jump process obtained in the limit of \(\Gamma\to0\).
    Circles represent discrete states labeled by \(n\) and \(i\), and arrows represent the transitions among them.
    The horizontal axis indicates the angular position of the minima of the potentials corresponding to the discrete states.
    }
    \label{f:6-states_discrete}
\end{figure}
The enzyme kinetics of the fine-grained model in the limit of \(\Gamma\to0\) can be calculated analytically.
In this limit, it is expected that the angle \(\theta\) rapidly equilibrates in each potential, and the dynamics of \(n\) and \(i\) follows a Markov jump process [Fig.~\ref{f:6-states_discrete}].
The transition rates from \((n,i)\) to \((m,j)\), \(R_{n,i}^{m,j}\), are evaluated as
\begin{widetext}
\begin{align}
 \begin{split}
     R_{n,\B}^{n+1,\h} &= \int \mathrm d\theta R_{n,\B}^+(\theta) P^\mathrm{eq}(\theta | n,\B) = W =: \Weff^+
     ,
     \\
     R_{n,\h}^{n-1,\B} &= \int \mathrm d\theta R_{n,\h}^-(\theta) P^\mathrm{eq}(\theta | n,\h) = W \exp\left[-\frac{1}{\kb T}\left(\Delta\mu - \tdm + \Nex \cdot \SI{80}{\degree}\right)\right] =: \Weff^-
     ,
     \\
     R_{n,\h}^{n,\B} &= \int \mathrm d\theta R_{n,\h}^+(\theta) P^\mathrm{eq}(\theta | n,\h) = W' \exp\left\{\frac{1}{\kb T}\left[-q'(1-q')\frac{k}{2}(\SI{40}{\degree})^2 + q' \left(\tdm + \Nex \cdot \SI{40}{\degree}\right)\right]\right\} =: \Weff'^+
     ,
     \\
     R_{n,\B}^{n,\h} &= \int \mathrm d\theta R_{n,\B}^-(\theta) P^\mathrm{eq}(\theta | n,\B) = W' \exp\left\{\frac{1}{\kb T}\left[-q'(1-q')\frac{k}{2}(\SI{40}{\degree})^2 - (1-q') \left(\tdm + \Nex \cdot \SI{40}{\degree}\right)\right]\right\} =: \Weff'^-
     ,
 \end{split}
\end{align}
\end{widetext}
and 0 otherwise.
Here, 
\begin{align}
    P^\mathrm{eq}(\theta| n, i) = \frac{\exp\left[-\frac{U_{n,i}(\theta)}{\kb T}\right]}{Z_{n,i}}
\end{align}
are the equilibrium angular distributions in \(U_{n,i}(\theta)\),
and \(Z_{n,i} = \int \mathrm d\theta \exp\left[-\frac{U_{n,i}(\theta)}{\kb T}\right]\) are the partition functions.
The rotational rate \(V\) is expressed in the form of the Michaelis-Menten equation as
\begin{align}
    V = \frac{1}{3} \frac{(\Weff^- + \Weff'^+)\Weff^+ - (\Weff^+ + \Weff'^-)\Weff^-}{\Weff^+ + \Weff^- + \Weff'^+ + \Weff'^-}
    = \frac{V_\mathrm{max} W}{K_\mathrm{m} + W}
    \label{e:low frictional limit}
\end{align}
with 
\begin{align}
    V_\mathrm{max} &= \frac{1}{3}\frac{\Weff'^+  - \Weff'^- \exp\left[-\frac{1}{\kb T}\left(\Delta\mu - \tdm + \Nex \cdot \SI{80}{\degree}\right)\right]}{1 + \exp\left[-\frac{1}{\kb T}\left(\Delta\mu - \tdm + \Nex \cdot \SI{80}{\degree}\right)\right]}
    ,
    \\
    \Km &= \frac{\Weff'^+ + \Weff'^-}{1 + \exp\left[-\frac{1}{\kb T}\left(\Delta\mu - \tdm + \Nex \cdot \SI{80}{\degree}\right)\right]}
    .
\end{align}
In Fig.~\ref{f:crossover}, it can be seen that the enzyme kinetics approaches to Eq.~(\ref{e:low frictional limit}) as \(\Gamma\) decreases.

\end{document}